# Dynamic Phase Diagram of an Orthogonal Spin Torque device: Topological Category

Yuan Hui, Zheng Yang and Hao Yu[a]

*Department of Physics, Xi'an Jiaotong-Liverpool University, 111 Ren'ai Rd., Suzhou 215123, P. R. China*

Abstract

The magnetization evolution of the free layer in an orthogonal spin-torque device is studied based on a macrospin model. The trajectory of magnetization vector under various conditions has shown rich nonlinear properties. The phase diagram is obtained in the parameter spaces of current density and the polarization distribution (the ratio of in-plane and out-of-plane polarizers). These dynamic phases can be classified according to their nonlinear behaviors which are topologically different, namely limit point and/or limit cycle. The topological classification is meaningful to design the ultra-fast spin-torque devices under different dynamic conditions towards various applications such as memory and oscillators.

Introduction

Spin-transfer torque (STT) discovered by Slonczewski and Berger [1-2] is an effect that conductive electrons carrying angular momentum reorient the local spins, which enables the manipulation of magnetization by a spin-polarized current flowing through a multi-layered junction. Current-induced STT expands the writing technique for data storage such as the magnetic random-access memory (MRAM) [3, 4]. It is also promising for microwave oscillators owing to the ultrafast processional switching of magnetic free layer [5]. A STT nanopillar magnetic tunnel junction (MTJ) consist of a free layer and one or two reference layer(s) where the magnetization of layers can be either in-plane or out-of-plane. The combination of orthogonal (out-of-plane) polarization free layer with in-plane reference layers (Fig. 1, (b)), firstly proposed by Kent [6], has been demonstrated to be able to achieve more efficient layer magnetization [7]. In such orthogonal spin torque (OST) device, the maximum spin-transfer torque from the beginning of the current pulse causes faster reversal of layer magnetization and less switching energy is cost, therefore it could be seen that the polarization comprising both in-plane polarizer and perpendicular one optimizes the original STT sandwich model and realizes a more ideal writing technique.

---

[a] Author to whom correspondence should be addressed. Email: hao.yu@xjtlu.edu.cn

The evolution dynamics of magnetization in OST device have been studied by previous researchers using analytical calculation and numerical simulation on various macrospin models [8-10]. They have shown that (i)the free layer magnetization can be ten times faster in OST than that of an in-plane-polarizer-only device [8]; (ii) steady precession can be excited by a current and the precession frequency depends on the strength of orthogonal polarizer[9]; (iii) time for complete reversal, from parallel (P) to anti-parallel (AP) or from AP to P, can be shortened by adjusting pulse width[10], and consequently the idea of half-precession was generated that can be achieved by controlling the pulse duration, in which case the switching energy is reduced as well. The current-field state diagram [11] for an OST device has been presented experimentally illustrating the range of different states. Low temperature OST memory element has been studied [12] over a wide range of parameter space and again demonstrated the switching dynamics are dominated by the out-of-plane spin polarization.

Previous research works have demonstrated that a macrospin model is effective to reflect the intrinsic property of the magnetization dynamics of an STT/OST device. However the dynamics of such spin oscillators are very complicated with rich physics in terms of its nonlinearity. In a conventional STT device (only with in-plane reference layer), the transition to chaotic dynamics has been revealed in a study on Landau-Lifshitz-Gilbert-Slonczewski (LLG) equation, showing a series of period doubling bifurcations[13]. The rich nonlinear phenomena suggest that we could classify the dynamic system in terms of its nonlinearity such as limit cycle/limit point formed by the evolution trajectory of magnetization vector. In fact, in this macrospin OST model, we discovered that the dynamic process of magnetization evolves with horizontal equilibria bifurcation under particular conditions of current density and polarization distribution (the ratio of out-of-plane polarizer to the in-plane one), which offers a reference to different applications. For each phase, a specific range of current density and polarization distribution are provided.

In this paper, we demonstrate the dynamic magnetization process by numerical simulations and discuss the results from the perspective of topological phase transition in which current density can causes bifurcations. The following contents will be orderly introducing the model, results in processional topology along with analysis of relationship between distribution of polarization and current density-induced phase transition, and final conclusion with related applications.

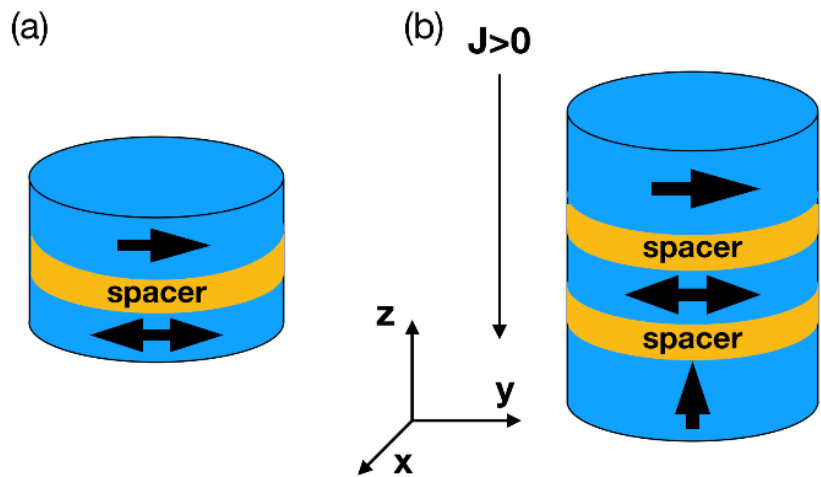

Fig. 1. Schematic of a cell of sandwich spin valve device: (a) a conventional STT nanopillar with in-plane polarizing magnetization, consisting of an analyzer and a free layer (from top to the bottom) ; (b) a orthogonal STT nanopillar with an out-of-plane polarizing reference layer in the bottom. The easy-axis of magnetization of free layer is along y direction.

We adopted a macrospin model from the reference [10], which was initially constructed according to Landau-Lifshitz-Gilbert-Slonczewski equation [1, 2, 14],

$$\frac{d\vec{m}}{dt} = -\gamma \vec{m} \times \vec{H}_{eff} + \alpha\gamma\vec{m} \times (\vec{m} \times \vec{H}_{eff}) - \tau_\parallel \vec{m} \times (\vec{m} \times \vec{n}_y) + \tau_\perp \vec{m} \times (\vec{m} \times \vec{n}_z)$$

(1)

$$\tau_\parallel = \gamma a_\parallel; \tau_\perp = \gamma a_\perp; \vec{H}_{eff} = -H_d m_z \vec{n}_z + H_K m_y \vec{n}_y$$

where $\vec{m} = \frac{\vec{M}_s}{M_s}$ is the normalized magnetization and $\vec{M}_s$ is the saturation magnetization of the free layer, $\vec{n}_y$ and $\vec{n}_z$ are the unit vector along y and z respectively, $\gamma = \gamma_0/(1+\alpha^2)$ where $\gamma_0$ is the gyromagnetic ratio and $\alpha$ is Gilbert damping parameter, $\vec{H}_{eff}$ is the effective field defined as $\vec{H}_{eff} = -H_d m_z \vec{n}_z + H_K m_y \vec{n}_y$ where $H_d$ is the demagnetizing field and $H_K$ is the uniaxial shape anisotropy field. The two coefficients $\tau_\parallel$ and $\tau_\perp$ are torques due to the in-plane and out-of-plane reference layer respectively and are defined [10] as $\tau_\parallel(\tau_\perp) = \frac{\hbar \eta_\parallel (\eta_\perp)}{2e\mu_0 M_s d_z} J$ where $e$ is electron charge, $\mu_0$ is permeability of vacuum and $\hbar$ is Plank constant, $d_z$ is thickness of free layer and $J$ is current density, $\eta_\parallel (\eta_\perp)$ is the current polarization of the parallel and perpendicular reference layer. The external current is provided as pulse which is downward along the z axis, shown in Fig. 1 (b). We took reference results as the final equilibrium polarized stage of free layer which is made of iron [15]. The size of the free layer is with length $d_y$ =100nm, width $d_x$=50nm and thickness $d_z$=5nm. More detailed value of parameters are defined in work from Mejdoubi et, al. (2013)[10]. The free layer is only affected by spin-transfer torque generated by two polarizers through current pulse. No more external field is considered.

The polarization distribution factor $r$ is defined as the ratio of the spin-torque amplitude of in-plane analyzer versus perpendicular polarizer [10]. When $r=0$, perpendicular polarizer dominants the polarization of free layer and meanwhile, $r=\infty$ means the major influence is from in-plane analyzer. Since this work is mainly focused on the dynamic evolution of the magnetization vector of free layer, we did simulations for each fixed $r$ by applying an external current pulse, with width $\delta=1$ ns. And we further analyzed different results correspondingly.

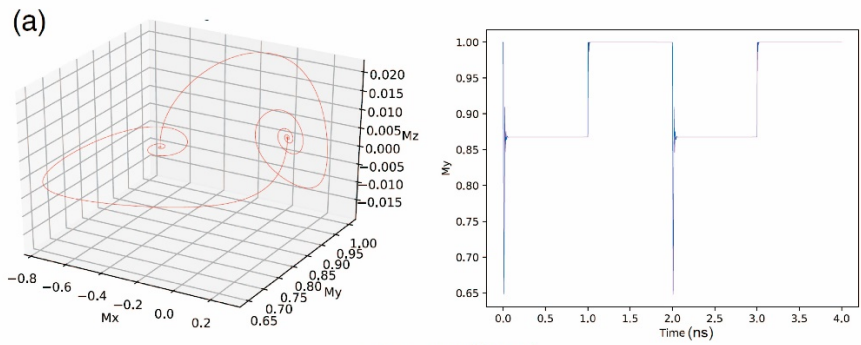

$r=1.0, J=3\times 10^{10} A/m^2$

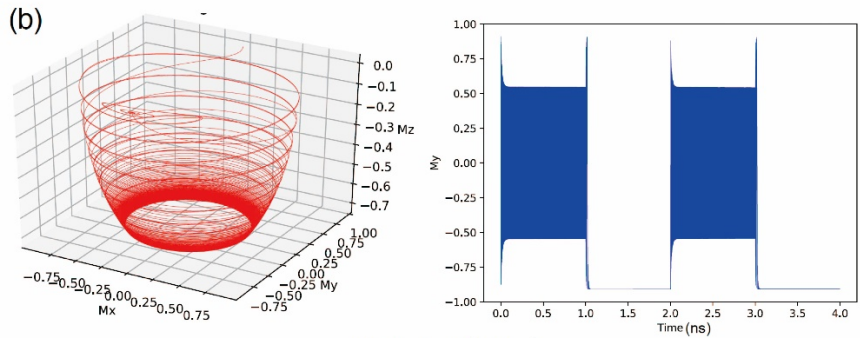

$r=1.0, J=3\times 10^{11} A/m^2$

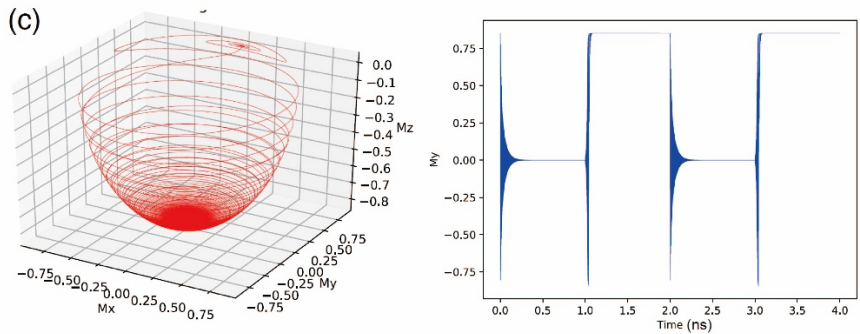

$r=1.0, J=4\times 10^{11} A/m^2$

Fig. 2.  Each phase of magnetization process with r=1 and (a) $J = 3 \times 10^{10} A/m^2$, (b) $J = 3 \times 10^{11} A/m^2$ and (c) $J = 4 \times 10^{11} A/m^2$ respectively. The diagram on the left: 3D magnetization vector (normalized $m_x$, $m_y$, $m_z$) evolution; right: magnetization along easy axis $m_y$ respond to current pulse in two periods. Nonlinear behaviors can be found from the left diagrams as (a) two limit points with different $m_x$, (b) one limit point and one limit circle, and (c) two limit points along z.

The spin torque is maximum at the beginning owning to the orthogonal polarization reference layer, therefore the tilting or switching occurs very fast after a while. As the result, Fig. 2 shows the phase transition as an example taken r=1 when current density increases ((a) $J = 3 \times 10^{10} A/m^2$, (b) $J = 3 \times 10^{11} A/m^2$ and(c) $J = 4 \times 10^{11} A/m^2$). The left side lists 3D diagram of normalized magnetization vector for each case and is attached with corresponding oscillation of $m_y$, the magnetization along easy axis of free layer, namely on the right side. Transition appears as current density increases. Specifically, Phase 1 (Fig.2, (a)) represents the condition when current density ($J = 3 \times 10^{10} A/m^2$) is below the critical one that is insufficient for complete switching. Phase 2 (Fig.2, (b)) reports the stage when current density is enough for magnetization ($J = 3 \times 10^{11} A/m^2$) where free layer could achieve a reversal of $m_y$, namely P to AP. Within the first half period 1ns, $m_y$ oscillates fast with the stable processional frequency [9]. There are two equilibrium states for Phase 2, one limit circle in the first half period and one limit point in the second half period. The stability of these equilibria changes when current density increases. It is topologically different from Phase 1 where the two states are both limit points. This would be discussed in the following contents. For the last stage, Phase 3 (Fig.2, (c)), when current density is large enough ($J = 4 \times 10^{11} A/m^2$), the middle oscillation converges as the limit circle transforms into another limit point from the point of view of topology.

Because of the contribution of the perpendicular polarizer, free layer magnetization vector oscillates before reach the final state. Therefore, the oscillation of free layer has different form in each situation, known from Fig.2. Particularly, appearance of well-defined frequency in Phase 2 relates to the system equilibria asymptotically approaching to the limit cycle.

In addition, each phase counters a range of current density and the limiting case causes bifurcation of the system [13, 16]. When $r$ is fixed, the main contribution of magnetization from in-plane analyzer or perpendicular polarizer is determined. Therefore, the critical current density to bifurcation could be obtained for each $r$. In our simulation, the fixed current width $\delta = 1ns$ promises following results, which are three different phases. However, changing the width of the current pulse, corresponding critical current density for phase transition could also be altered [10]. The following explanation and the phase diagram (Fig. 3) are about the relationship of current density and polarization distribution.

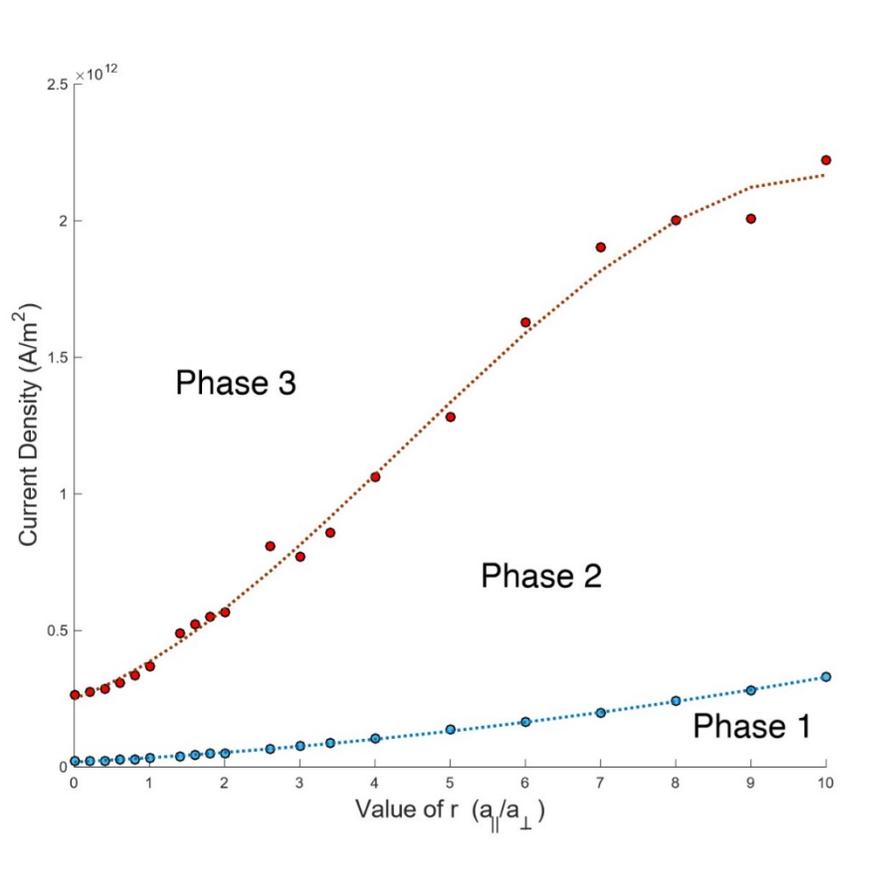

Fig. 3.  Phase diagram shows the critical current density for each case with different *r*. For *r* in range of [0.0,10], the system transforms through all three phases. As *r* keeps increasing, available range of current density to enable the system to reach Phase 2 enlarges, where well-defined frequency appears.

As mentioned in previous work [10], the frequency of oscillation in $m_y$ decreases as *r* becomes larger. This is because for a certain value of current density, the value of *r* determines how much torque can be transferred from in-plane analyzer and perpendicular polarizer. Hence, when *r* increases, the horizontal polarization dominates, which weakens the influence of perpendicular polarization so that there is less oscillations. Since our study shows that for each condition with respect to different *r*, J varies as well in case to cause the complete polarization of free layer, we restricted situation to only Phase 2, where the stable frequency would appear to study the further result in frequency and relationship between J and frequency.

Fig. 4. explains the oscillation frequency in y-plane in different cases of *r* with respect to *J* to reach Phase 2. It could be seen that when *r* becomes larger, the corresponding range of current density for Phase 2 expands. Therefore, we did not study the frequency for a fixed current density but as a function *f(J)*, where we simulated in Phase 2 condition.

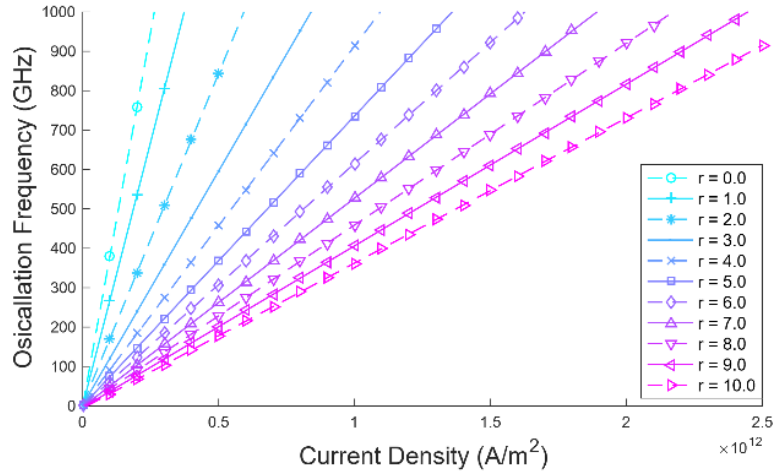

Fig. 4. Frequency with corresponding Phase 2 current density for each *r*. The fit lines illustrates the linear relationship between current density and oscillation frequency in y-plane.

The results show that *f(J)* meets the linear relationship in each group of frequency. Combined with the aforementioned studies as analysis for extreme scenario, when r keeps increasing to infinity, the required current density to cause reversal goes to infinity as well. Then Phase 2 disappears where no regular oscillation in y-plane could be obtained.

In conclusion, our study is based on the model of spin-transfer-torque MTJ with additional perpendicular polarizer. The results show that due to different current density and the ratio of in-plane polarizer to perpendicular polarizer, the dynamic features of magnetization evolution trajectories vary in terms of topology, which are summarized as follows: (1) two limit points for Phase 1 where current density below the critical value and no switching occurs; (2) one limit point and one limit cycle for Phase 2 with current density greater than the critical point, where the magnetization switches fast even within one pulse; (3) two limit points for Phase 3 with even larger current, the system switches between two limit points. The category of devices according to the topology of nonlinear dynamics can understand the STT-MTJ devices from a new perspective, and also offers a reference to design devices for various application purposes, such as ultra-fast microwave oscillators or MRAM. One can adjust the parameters of devices, namely *r* and *J*, to make their value fall in the appropriate phase and topological category.

**ACKNOWLEDGMENTS**

This research was supported by Key Programme Special Fund (KSF-E-22) and Research Enhancement Fund (REF17-1-7) of Xi'an Jiaotong-Liverpool University.